\documentclass[conference]{IEEEtran}
\usepackage{graphicx}
\usepackage{listings}
\usepackage{hhline}
\usepackage{amsmath}
\usepackage{amssymb}

\usepackage[usenames,dvipsnames]{xcolor}
\lstdefinestyle{SMV} {language=C,captionpos=b,tabsize=3,frame=lines,keywordstyle=\color[rgb]{0.6,0,0},morekeywords={MODULE, VAR, ASSIGN, SPEC, init, next, case,esac,IVAR,self,INVARSPEC,CTLSPEC},numbers=left,showtabs=false,morecomment=[l]--,commentstyle=\color{gray},
breaklines=true,stringstyle=\color[rgb]{0.627,0.126,0.941},
basicstyle=\footnotesize\ttfamily}
\lstdefinestyle{SSI} {language=C,captionpos=b,tabsize=3,frame=lines,keywordstyle=\color[rgb]{0.6,0,0},morekeywords={or, then, s, End, c, cr, cn, cfr, cfn, xs, l, f, clear, bpull },numbers=left,showtabs=false,morecomment=[l]//, commentstyle=\color{gray},
breaklines=true,stringstyle=\color[rgb]{0.627,0.126,0.941},
basicstyle=\footnotesize\ttfamily}
\lstdefinestyle{SCA} {language=C,captionpos=b,tabsize=3,frame=lines,keywordstyle=\color[rgb]{0.6,0,0},morekeywords={var, val, foreach, for, match},numbers=left,showtabs=false,morecomment=[l]//,commentstyle=\color{gray},
breaklines=true,stringstyle=\color[rgb]{0.627,0.126,0.941},
basicstyle=\footnotesize\ttfamily}
\lstdefinestyle{XML} {language=xml,captionpos=b,tabsize=3,frame=lines,keywordstyle=\color[rgb]{0.6,0,0},morekeywords={signal},numbers=left,showtabs=false,morecomment=[l]//,commentstyle=\color{gray},
breaklines=true,stringstyle=\color{gray},
basicstyle=\footnotesize\ttfamily}
\usepackage{enumerate}
\usepackage{textcomp}

\usepackage{url}
\usepackage[ruled,vlined,linesnumbered]{algorithm2e}
\IEEEpubid{}

\title{Verification of Interlocking Systems Using Statistical Model Checking}

\author{\IEEEauthorblockN{Quentin Cappart\IEEEauthorrefmark{1}, Christophe Limbr\'{e}e\IEEEauthorrefmark{1}, Pierre Schaus\IEEEauthorrefmark{1}, Jean Quilbeuf\IEEEauthorrefmark{2}, Louis-Marie Traonouez\IEEEauthorrefmark{2} and Axel Legay\IEEEauthorrefmark{2}}
\IEEEauthorblockA{\IEEEauthorrefmark{1}Universit\'{e} catholique de Louvain, Louvain-La-Neuve, Belgium \\
Email: \{quentin.cappart | christophe.limbree | pierre.schaus\}@uclouvain.be}
\IEEEauthorblockA{\IEEEauthorrefmark{2}INRIA/IRISA, Rennes, France\\
Email: \{jean.quilbeuf | louis-marie.traonouez | axel.legay\}@inria.fr}}

\begin{document}

%

\setlength{\IEEEilabelindent}{\IEEEilabelindentA}

\maketitle


\begin{abstract}
\footnote{This paper is published at HASE 2017 \cite{7911872}.}In the railway domain, an interlocking is the system ensuring safe train traffic inside a station by controlling its active elements such as the signals or points.
Modern interlockings are configured using particular data, called application data, reflecting the track layout and defining the actions that the interlocking can take. The safety of the train traffic relies thereby on application data correctness, errors inside them can cause safety issues such as derailments or collisions.
Given the high level of safety required by such a system, its verification is a critical concern. In addition to the safety, an interlocking must also ensure that availability properties, stating that no train would be stopped forever in a station, are satisfied. Most of the research dealing with this verification relies on model checking. However, due to the state space explosion problem, this approach does not scale for large stations.
More recently, a discrete event simulation approach limiting the verification to a set of likely scenarios, was proposed. The simulation enables the verification of larger stations, but with no proof that all the interesting scenarios are covered by the simulation.
In this paper, we apply an intermediate statistical model checking approach, offering both the advantages of model checking and simulation. Even if exhaustiveness is not obtained, statistical model checking evaluates with a parametrizable confidence the reliability and the availability of the entire system. 

\end{abstract}

\section{Introduction}
An interlocking is a system that controls the train traffic by acting as an interface between the trains and the railway track components. The track components are for example, the signals that allow the train to proceed, or the points that guide the trains from one track to another. The paths followed by the trains are called routes. Modern interlockings are computerized systems that are composed of generic software and application data. The generic software implements the signalling principles that are applicable to the railway domain. The development and the validation of that generic software follow the highest safety standards applicable to the domain. On the other hand the application data are specific to each station controlled by the interlocking. The ability of the interlocking to avoid critical situations, like train collisions, relies on the safety level achieved by the combination of the generic software and of the application data. Beyond the safety, an interlocking must also ensure that no train will be stopped too long in the station in order to maintain the availability of the station.
It is why availability properties must also be considered.
Most of the time, the validation of the application data is performed through testing on a physical simulator that reproduces the environment of the interlocking. This process is costly and error-prone. Moreover manual testing does not always cover all the scenarios that could possibly end up in an unsafe situation. \\

Until now, most of the research targeting the verification of the application data is based on model checking (\cite{cimatti1998formal,fantechi2012distributing,haxthausen2013applied}) even if other approaches based on formal methods exist \cite{fantechi2013some}. First, the signalling principles and the application data are translated into a model. Secondly, the dangerous situations that the interlocking must avoid are translated into safety properties. Finally, a model checker tool unrolls the state space of the model and verifies that none of the reachable states violates the properties. Following this approach, Busard et al. \cite{busard2015verification} designed a verification model for Belgian interlockings. The principle of verification is simple but suffers from the so-called state space explosion problem. In other words, the number of states is so large that its creation and exploration take exponential time. Different methods have been studied to limit this problem. Based on model checking, Winter and al. \cite{winter2006tool} proposed to relax the verification process by reducing the complexity and the size of the model. Furthermore, different optimisations and new algorithms (\cite{eisner1999using,huber2002towards,winter2012optimising}) aiming to accelerate the verification were also proposed. \\

A different approach, introduced by Cappart et al. \cite{cappart_simu} consists of performing the verification by a discrete event simulation.
The idea is to simulate the behaviour of an interlocking as described in its application data and to observe if any unwanted scenario occurs. 
Unlike model checking where all the states are considered, this approach mainly considers scenarios that are likely to happen in practice.
However, simulation provides no guarantee whatsoever that all such scenarios will be detected. \\

In this paper, we propose a method to automatically verify an interlocking using simulation and statistical model checking. The contributions of this paper are as follows:

\begin{itemize}
\item[$\bullet$] The introduction of two availability properties that an interlocking must face in order to ensure that no train will be stuck in a station. Concretely, we verify that each route can be released after being set and that no component is locked forever.
\item[$\bullet$] The formalisation of availability and safety properties of \cite{cappart_simu} in a bounded linear temporal logic (BLTL) \cite{kamide2012bounded}. Such a logic is used in order to bound the simulation time.
\item[$\bullet$] An extension of the simulation model described in \cite{cappart_simu} so that it can be automatically built from the application data and the topology of a station. The model has also been extended in order to support its verification for properties expressed in a bounded linear temporal logic and using statistical model checking algorithms.
\item[$\bullet$] The use of statistical model checking \cite{legay2010statistical} algorithms such as Monte Carlo, Chernoff's bound and importance splitting, for verifying this model. Using both statistical model checking and BLTL, we can now detect errors that could not be detected in \cite{cappart_simu}. These errors are related to availability properties. More generally, verification of any properties expressed in BLTL can now be performed, whereas the simulator of Cappart et al. could only support verification of invariants.
\end{itemize}


This paper uses a typical medium-sized Belgian station as a case study.  
The next section introduces the case study and describes how it is managed by an interlocking.
In Section 3, we explain how the problem is modelled.
Its verification is discussed in Section 4.
Finally, in Section 5, we analyse the performances of the verification and its reliability through experimental results.   

\section{Interlocking principles}

\begin{figure*}
\centering
\includegraphics[width=1\linewidth]{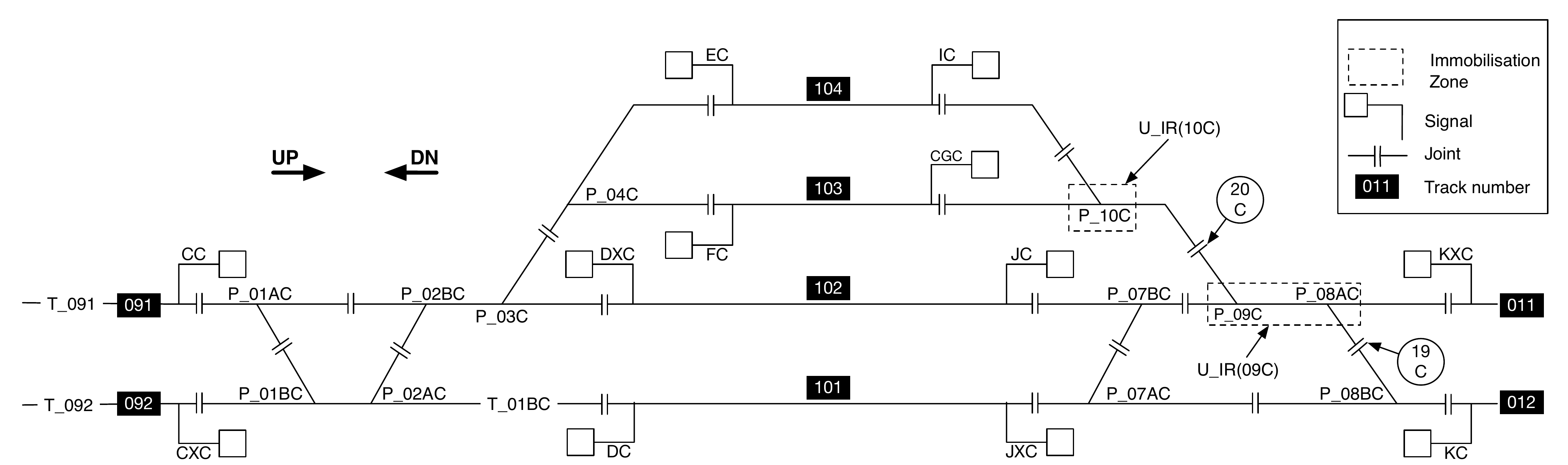} 
\caption{Layout of Braine l'Alleud Station}
\label{fig:braine} 
\end{figure*}

The case study targets Braine l'Alleud Station, a medium-sized railway station of the Belgian network. 
A representation of its track layout is shown in Figure \ref{fig:braine} (the names of some components are not written).
We consider the following physical components:

\begin{itemize}
\item[$\bullet$] The \textbf{points} (e.g. P\_01BC) are the track components which guide the train from one track to another. 
According to the Belgian convention, they can be in a normal position (left) or in a reverse position (right).

\item[$\bullet$]  The \textbf{signals} (e.g CC)  are the interface between the interlocking and the trains. 
They display a proceed state (green) when it is safe for a train to proceed on the station. 

\item[$\bullet$]  The \textbf{track segments} (e.g T\_01BC) are the tracks where a train can be detected. They can be either occupied by a train or clear.
They are delimited by the \textbf{joints}.
\end{itemize}

Besides the physical components, an interlocking also uses the notion of route.
A route is the path that a train must follow inside a station. A route is named according to its origin and destination point.
For instance, Route R\_CC\_102 starts from Signal CC and ends on Track 102.
When a train is approaching a station, a signalman performs a route request to the interlocking in order to ask if the route can be commanded for this train.
The interlocking handles this request and will accept or reject it according to the station state.
To do so, an interlocking uses logical components like the subroutes or the immobilisation zones, materialising the availability of some physical components.
Such components are locked or released if they are not requested.
Braine l'Alleud Station is controlled by a unique interlocking composed of 32 routes, 12 signals, 17 track segments, and 12 points. 

\section{Model of an interlocking system}
In this section, we present how we model an interlocking.
The objective pursued is to build a tool that automatically verifies the application data correctness of an interlocking system. To do so, our tool requires three inputs: 

\begin{itemize}
\item[$\bullet$] The \textbf{application data}, that describes the behaviour of the interlocking system.   

\item[$\bullet$]  The  \textbf{track layout}, that describes the geographic disposition of the physical components in the station.  

\item[$\bullet$]  The properties that the interlocking must satisfy. They are twofold: on the one hand, the \textbf{safety properties}, which state that no issue such as a collision or a derailment occurs in the station, and on the other hand, the \textbf{availability properties} that forbid the interlocking to block trains for too long. 

\end{itemize}
From these inputs, our tool provides statistics about the correctness of the model according to the properties defined. 
The verification process is divided into four steps:

\begin{enumerate}

\item  Generating an interlocking model combining the application data and the track layout of the related station. This is done using two translators that parse and aggregate both data sources into a single model. 

\item  Stating all the safety and availability properties that must be satisfied. Properties have been formalised in BLTL. This logic requires that linear temporal operators have bounds. These bounds guarantee that a property can be decided on a finite length simulation.

\item  Simulating the behaviour of the obtained model with a discrete event simulator. A set of traces summarizing the different actions that occurred during the simulation is obtained through this process. The simulation is based on the verification tool introduced and developed by Cappart et al. \cite{cappart_simu}.

\item Evaluating the probability that the simulations satisfy the properties. To this end, we use a statistical model checker. This tool has been implemented using PLASMA Lab \cite{boyer2013plasma} which is a platform for statistical model checking of stochastic models.  
\end{enumerate}

\begin{figure}
\centering
\includegraphics[width=\linewidth]{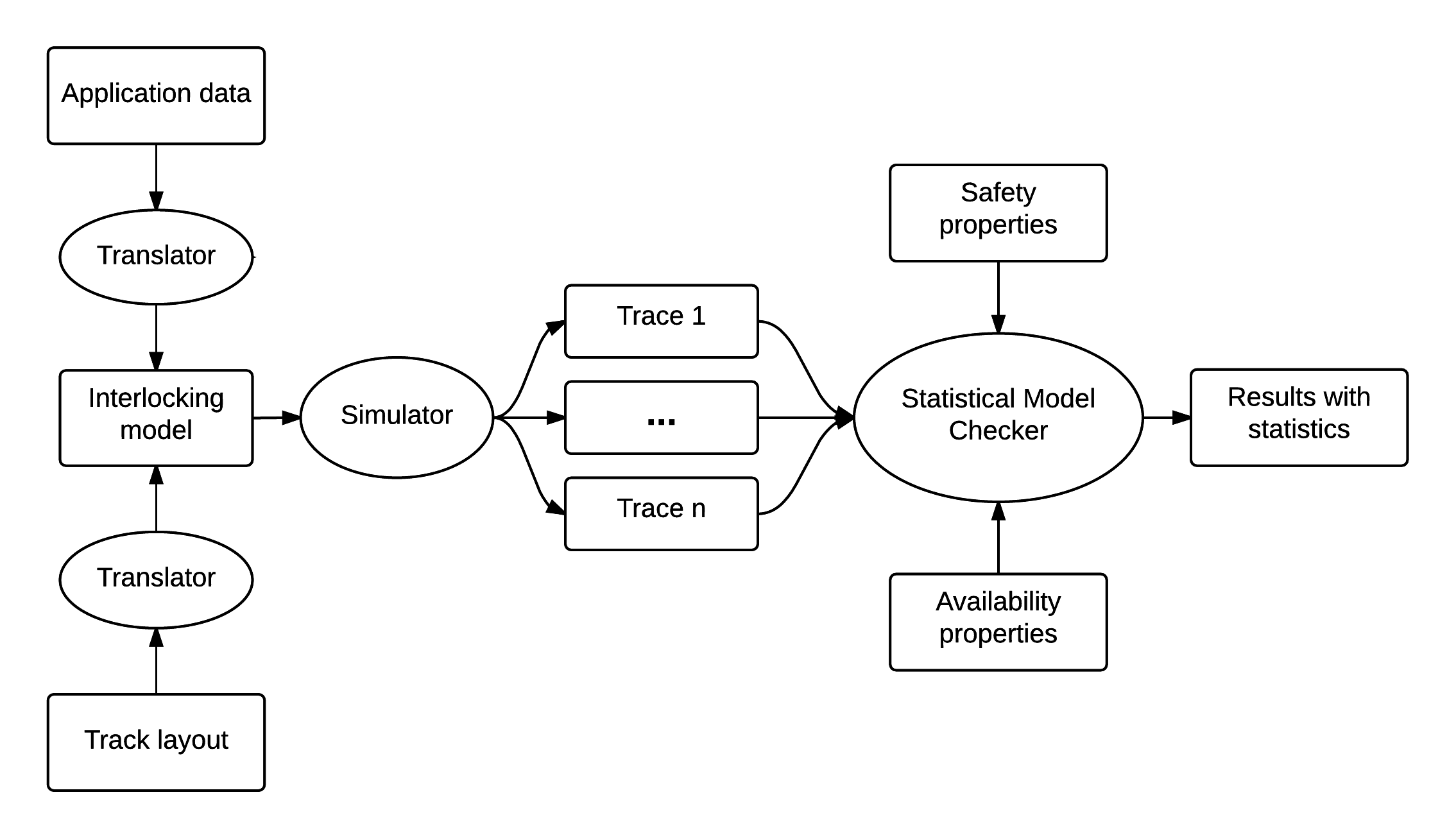} 
\caption{Steps of our approach}
\label{fig:process} 
\end{figure}

The data flow diagram presented in Figure \ref{fig:process} resumes the approach. 
The whole process is entirely automatised.
Next subsections describe in more details the different steps.

\subsection{Translation of the input data}

The first step is to build from the input data a model reflecting the behaviour of an interlocking applied to its station.
Two data types are involved in this process, the application data and the track layout. 
\\

The format considered for the application data is Solid State Interlocking (SSI) \cite{cribbens1987solid}, used by the Belgian railways since 1992. The methodology described in this paper can nevertheless be applied to other railway formats. 
The application data describe the behaviour of an interlocking. The actions that the interlocking can perform and the conditions under which they can be executed are described inside. To do so, two logical components are used: 

\begin{itemize}
\item[$\bullet$] The \textbf{subroutes}: they represent the contiguous segments that the trains must follow inside a route. When a route is commanded for a train, a set of subroutes is locked. When not requested, subroutes are in a free state. In application data, they are defined by this syntax: U\_\textit{origin}\_\textit{destination}.
\item[$\bullet$] The \textbf{immobilisation zones}:  they are the variables materialising the immobilisation of a set of points. When they are locked, their attached points cannot be moved. They are represented in the application data by the name U\_IR.
\end{itemize}

Other components are also present in the application data but they are either not related to the safety/availability or abstracted in our model.
Using such components, an interlocking can control the train traffic by monitoring the station, commanding routes, locking components and releasing them. All the possible actions and their underlying conditions are described in the application data. For instance, they contain the following instructions:

\begin{itemize}
\item[$\bullet$] The \textbf{route requests}: they determine the conditions that must be satisfied before commanding a route and the actions that the interlocking must take to fulfil the request. Listing \ref{routeRequest} shows the definition of a request for Route R\_CGC\_011 from Signal CGC to Track 011. This route can only be set if it is not already set (xs on line 2), if some points are free to be commanded to the reverse (cfr) or the normal (cfn) position (lines 3-4) and if some immobilisation zones are free (f) to be used (line 5). If all of these conditions are satisfied, the request is granted and
the route is set (s on line 6), the points are commanded (cr and cn) to the requested positions (lines 7-8) and the immobilisation zones (line 9) and some subroutes (line 10) are locked (l). 

\begin{lstlisting}[style=SSI, caption= Request for setting route R\_CGC\_011,  label=routeRequest]
*Q_R(CGC_011)
	if		R_CGC_011 xs 
			P_09C cfr, 
			P_10C cfn, P_08AC cfn, P_08BC cfn
			U_IR(10C) f, U_IR(09C) f
	then	R_CGC_011 s
			P_09C cr, 
			P_10C cn, P_08AC cn, P_08BC cn, 
			U_IR(10C) l, U_IR(09C) l, 
			U_CGC_20C l, U_20C_KXC l
\end{lstlisting}

\item[$\bullet$] The  \textbf{point commands}: they determine the conditions that must be satisfied before commanding a point. Listing \ref{pointRequest} shows an instantiation of these rules for Point P\_09C. Typically, a point requires its immobilisation zone to be free (f) for its command. 

\begin{lstlisting}[style=SSI, caption= Request for commanding Point P\_09C,  label= pointRequest]
*P_09CN U_IR(09C) f // condition for normal (N) position
*P_09CR U_IR(09C) f // condition for reverse (R) position
\end{lstlisting}

\item[$\bullet$] The \textbf{releasing requests}:  they determine the conditions that must be satisfied to release a component. Listing \ref{compFreeing} states that Subroute U\_CGC\_20C can
only be released if Routes R\_CGC\_011 and R\_CGC\_012 are not already set (xs) and if there is no train (c) on Track segment T\_10C. Besides,  Listing \ref{compFreeing_UIR} states that Immobilisation Zone U\_IR(09C) can be released only if some subroutes are free (f).

\begin{lstlisting}[style=SSI, caption= Request for releasing subroute U\_CGC\_20C,  label= compFreeing]
U_CGC_20C f if R_CGC_011 xs, R_CGC_012 xs, 
               T_10C c 
\end{lstlisting}

\begin{lstlisting}[style=SSI, caption= Request for releasing immobilisation zone  U\_IR(09C),  label= compFreeing_UIR]
*sub_free_09C
	U_18C_KXC f, U_KXC_18C f, U_18C_19C f, 
	U_19C_18C f, U_20C_19C f, U_19C_20C f,
	U_20C_KXC f, U_KXC_20C f
\end{lstlisting}

\end{itemize}

Such data describe the interlocking behaviour but contain no information about the track layout of the considered station.
However, the correctness of an interlocking is also dependent of its consistency with the track layout.
It is why a data source describing the track layout is necessary. \\

Based on an XML structure, the railway description language called railML \cite{nash2004railml} provides a universal format that can be used for diverse railway applications.  Several schemas such as the infrastructure, the timetables or the rolling stocks are described inside. In our case, only the infrastructure schema is necessary. \\
%
%

Our verification model is built from both SSI and railML languages.
It is done by two translators which automatically parse both data and aggregate them.
We use a graph structure, implemented in Scala programming Language \cite{odersky2004overview}, to model the station by using topology information present in the railML file.
Physical components such as the points, the signals or the joints are represented by nodes and the portions of track segments between them
correspond to the edges. Each edge belongs to a particular track segment. Trains move in the station from edge to edge. Once the graph is built, application data are then used to simulate the interlocking behaviour on this graph. Detailed explanations on the simulator are provided in the next section.

\subsection{Simulation of the model}

Once the model is obtained, the next step is to verify its correctness.
One common approach for that is model checking. However, 
because of the state space explosion problem, model checking does not scale for large stations.
To overcome this issue, Cappart et al. \cite{cappart_simu} introduced a verification method based on a discrete event simulation.
The key idea is to reproduce the interlocking behaviour under a realistic train traffic. 
If no issue occurred during simulations covering enough time, we can have a high expectation that the system is safe.
The main difference with model checking is that there is no guarantee that all the possible cases are considered but only the cases that can potentially happen in real situations. \\

The principles of our simulation is similar to the simulation introduced in \cite{cappart_simu}. 
Two kinds of entities are involved: On the one hand the trains, that are characterized by a direction and a position (a track segment)
and on the other hand the interlocking active components, as described in the previous section.
Besides, entity states change according to the following events: \\

\begin{itemize}
\item[$\bullet$] The \textbf{train arrivals}: trains arrive randomly at a particular start signal. In our simulation, a train arrival can occur with a uniform probability on the interval $[t_a, t_a +\beta_a]$ where $t_a$ is the time of the last train arrival ($t_a = 0$ for the first step) and $\beta_a$ is a predefined parameter. Besides, after each event occurrence, a new event is triggered in the interval $[t_a , t_a + \beta_a]$ with the updated $t_a$. \\

\item[$\bullet$] The \textbf{route requests}: route requests are performed for trains waiting at a start signal. The request is accepted only if all of its conditions are fulfilled. Otherwise, the request is aborted. In case of acceptance, the route is set and all the actions described in the request are executed. Otherwise, no action is taken. A route request is an event which can occur in the interval $[t_r,t_r+\beta_r]$ with $t_r$ the time of the last request and $\beta_r$ a predefined parameter. \\

\item[$\bullet$] The \textbf{train movements}: trains move through the station from track to track by following the direction set by the signals and the points. When a train reaches the end of its route, it is removed from the station. The first movement of a train $x$ is triggered when its route is accepted. The next movements occur in the interval $[t_m(x), t_m(x)+\beta_m]$ with $t_m(x)$ the time of the last movement done by $x$ and $\beta_m$ a parameter. Each train has thereby its own queue of events. By doing this, we implicitly model the fact that the speed of the trains can be different. \\

\item[$\bullet$] The \textbf{component releasing}: after each event, the interlocking verifies if components can be released. For instance, if the conditions presented in Listing \ref{compFreeing} are fulfilled, Subroute U\_CGC\_20C will be released.  \\
\end{itemize}

Randomness is introduced for several events through the parameters $\beta$. Their purpose is to allow the generation of different scenarios at each simulation by defining a time range on which the events can occur. Discussion about the choice of this parameter is presented in \cite{cappart_simu}. \\

A simulation of $n$ steps provides a trace of $n$ states.
A simulation state $s_i$ with $i \in [1,n]$ is defined as

\[ s_i: \Big<  nb, \sigma_p, \sigma_r, \sigma_s, \sigma_u, \sigma_t, \sigma_{tr} \Big>. \]

\begin{itemize}
\item[$\bullet$] $nb$ is the number of trains that have moved in the station so far. This variable is used to under approximate how many real
days the simulation has covered. Indeed, by taking the extreme case of a busy station where there is an incoming train every minute all the day long, we can
safely assume that the simulation has covered at least one real day when 1440 trains have moved through the station.

\item[$\bullet$] $\sigma_p: \texttt{point} \rightarrow \{\texttt{normal},\texttt{reverse}\}$ is a function defining the position of a point. 

\item[$\bullet$] $\sigma_r: \texttt{route} \rightarrow \{\texttt{set},\texttt{unset}\}$ is a function defining if a route is set or unset. 

\item[$\bullet$] $\sigma_s: \texttt{subroute} \rightarrow \{\texttt{free},\texttt{locked}\}$ is a function defining if a subroute is free or locked. 

\item[$\bullet$]  $\sigma_u: \texttt{uir} \rightarrow \{\texttt{free},\texttt{locked}\}$ is a function defining if an immobilisation zone is free or locked. 

\item[$\bullet$] $\sigma_t: \texttt{track} \rightarrow \mathbb{N}$ a function defining the number of trains being on a track segment. 

\item[$\bullet$] $\sigma_{tr}: \texttt{train} \rightarrow (\texttt{track},\{\texttt{up},\texttt{down}\})$ a function defining the current position of a train and its direction.

\item[$\bullet$] $\texttt{point}$, $\texttt{route}$, $\texttt{subroute}$,$\texttt{uir}$, $\texttt{track}$ 
 are the set of the interlocking components defined in the application data and $\texttt{train}$ the set of trains in the station.  
 
\end{itemize}

The simulator has been implemented using the discrete event simulation package of OscaR \cite{oscar}, a Scala toolkit for solving operations research problems.
Furthermore, advanced features have been added to the simulator engine in order to use the algorithms presented in the next section.
For instance, we added the possibility to save a simulation state and use it as an initial state for new simulations.

\subsection{Definition of properties}
\label{propDefSec}
Before verifying that no unwanted scenario occurs, we need to define what  is exactly an unwanted scenario.
As previously stated, the goal of an interlocking is to ensure a safe train traffic in a station.
To achieve this goal, an interlocking has two requirements.
On the one hand, it must ensure that it will cause no accident in the station. It is a \textbf{safety requirement}.
Busard et al. \cite{busard2015verification} identified three safety requirements:

\begin{itemize}
\item[(1)]  A track cannot have two trains on it at the same time in order to avoid collisions.
\item[(2)]  A point cannot move if there is a train on it otherwise it will derail.
\item[(3)]  A point must always be set on a position allowing trains to continue their path. Otherwise, the trains will derail.
\end{itemize}

On the other hand, an interlocking should not block trains or components for too long. It is an \textbf{availability requirement}. It can be refined into two requirements:

\begin{itemize}
\item[(4)]  A route could always be \textit{eventually} set.
\item[(5)]  Component could always be \textit{eventually} released.
\end{itemize}

\textit{Eventually} means that after any state of the system, there exists at least another state where the property is satisfied. These properties are expressed by mean of BLTL formulas. The following formulas show an instantiation of these properties in BLTL:

\begin{itemize}
\item[(1)] $ \textbf{G}_{n} \ \  \texttt{T\_01BC} \leq 1 $
\item[(2)]  $\textbf{G}_{n} \ \   \texttt{T\_01BC} = 1  \implies  \texttt{P\_02AC} = \textbf{next}(\texttt{P\_02AC}) $
\item[(3)] $\textbf{G}_{n} \ \  \big(\texttt{T\_092} = 1 \land \texttt{T\_01BC} = 0   \\  
\ \ \ \ \ \ \land  \textbf{next}(\texttt{T\_092}) = 0 \land \textbf{next}(\texttt{T\_01BC}) = 1\big) \\
 \ \ \ \ \ \ \implies \big(\texttt{P\_01BC} = \texttt{left} \land \textbf{next}(\texttt{P\_01BC}) = \texttt{left} \big)$
\item[(4)]  $(\textbf{GF})_{n} \ \   \texttt{R\_CC\_101} = \texttt{set}$
\item[(5)] $(\textbf{GF})_{n} \ \  \texttt{U\_CGC\_20C} = \texttt{free} $
\end{itemize}

Formula (1) states that for a simulation of $n$ steps, at any moment (\textbf{G}), at most one train can be on Track \texttt{T\_01BC}. Formula  (2) states
that if there is a train on Track \texttt{T\_01BC}, the direction on Point \texttt{P\_02AC} cannot change at the next state. Formula  (3) states that if there is a train moving from Track \texttt{T\_092} to Track  \texttt{T\_01BC}, Point \texttt{P\_01BC} must be set and stay to \texttt{left}. Such equations are related to the safety, the next ones ensure the availability of the system. Formula  (4) ensures that after any state, there exists at least another state where Route \texttt{R\_CC\_101} could be set.  Formula (5) has the same idea by ensuring that Subroute \texttt{U\_CGC\_20C} could always (\textbf{GF}), be released. Such equations illustrate the properties that the model has to satisfy. There are similar formulas for each component of the station. Formulas (1), (2) and (3) are expressed only in terms of physical components and can then be  automatically generated from the track layout. Concerning Formulas (4) and (5), they can be automatically generated from the set of the routes and subroutes involved in the application data.
\\

Such a formalism is generally used when a finite time domain is considered.
In our case, a BLTL property of bound $n$ is satisfied when there is no state that
violates the property during the $n$ first steps of the simulation. The property can thus be decided after at most $n$ steps of simulation. 
The statistical reliability of the verification by simulation is then highly dependent on the choice of this bound.
Indeed, if the bound is too low, the simulation time will be too short and some scenarios will not be covered.
It is why we require that the bound must be sufficient to determine a simulation time long enough to cover at least one \textbf{complete scenario}.
A complete scenario is a scenario going from a train arrival to its departure in a station. 
The scenario is not complete if the simulation is stopped when the train is still waiting or moving into the station.
For instance, a simulation of 1 hour will not be sufficient because trains arriving in a station could still be in the 
station after one hour. In our case, we assigned a simulation time of one complete day. 
This value is chosen under the reasonable assumption that a train would not stay into the same station longer than one day. With such a bound, we can have the certainty that all the situations have a non zero probability to occur during a simulation and that all the possible scenarios can be covered through a simulation. BLTL provides two ways of specifying the bounds of temporal operators. The first one uses the number of steps while the second one relies on a time unit defined in the model. In our case, we use the number $nb$ defined in the state of our model as the time unit. As explained previously, using a bound of 1440 ensures that our simulation covers at least a complete day.

\section{Verification by statistical model checking}

Now that the interlocking requirements have been stated, the next step is to evaluate whether the interlocking satisfies them.
The key idea is to perform several simulations, get the resulting traces, analyse them and verify that they contain no state violating the requirements.
A \textbf{statistical model checker} \cite{legay2010statistical} can be used for that. The aim of statistical model checking is to approximate, in a controlled manner, the probability of satisfaction or violation of a property.
Unlike classical model checking approaches where an exhaustive exploration of the state space is conducted (Busard et al. \cite{busard2015verification} experimented the limitation of this approach for SSI language), statistical model checking only requires a sample of simulations. 
On this section, we describe the statistical model checking algorithms used in our approach. \\

\subsubsection*{Monte Carlo estimation} The first algorithm is based on Monte Carlo method for estimating the probability $\gamma$ of satisfying a property $\varphi$.
The principle is to generate $N$ random simulations $\rho_1, \ldots, \rho_N$ and to compute the following estimation of $\gamma$:
\begin{equation}
\tilde{\gamma}=\frac{1}{N}\sum_{i=1}^{N} \mathbf{1}(\rho_i \models \varphi)
\end{equation}
where $ \mathbf{1}$ is an indicator function that returns $1$ if $\varphi$ is satisfied in $\rho_i$ and $0$ otherwise. \\

Parameter $N$ can be determined by the user according to the number of simulation he wants to perform. 
Given that property $\varphi$ has bounded temporal operators, a simulation will stop whenever the bound ($n$), or a state violating $\varphi$, is reached. \\

The parameter $N$ can also be determined in order to obtain a specific confidence on the probability obtained. 
For instance, the Chernoff's bound \cite{chernoff1952measure} determines the required number of simulations to perform in order to have a confidence $\delta$ and a precision $\epsilon$ on the value $p$ obtained:

\begin{equation}
\label{chernoff}
Pr(|\gamma - \tilde{\gamma}| < \epsilon) \geq 1 - \delta \quad \textrm{ if } \quad N \geq \frac{ln(\frac{2}{\delta})}{2\epsilon^2}
\end{equation}

This bound $N$ guarantees that the probability that a property is satisfied is included in the  $(1-\delta)$-$[\gamma-\epsilon, \gamma+\epsilon]$ confidence interval. \\



\subsubsection*{Importance splitting} Importance splitting \cite{JLS13}, a technique usually used for rare event detection, allows to increase the probability of generating rare events and to speed up the errors detection by decreasing the number of simulations required to estimate the probability.
\\

Importance splitting starts by splitting the rare property in a sequence of temporal properties $\varphi_{k}$, with the
logical characteristic
$
\varphi=\varphi_{M}\Rightarrow\varphi_{M-1}\Rightarrow\cdots\Rightarrow\varphi_{1}.
$
It defines a set of levels, each associated to the conditional probability $Pr(\rho \models \varphi_{k+1} \mid \rho \models \varphi_{k})$
of reaching level $k+1$ from level $k$.   Instead of trying to verify directly a rare property, the importance splitting algorithm considers a set of sub-properties easier to verify and which lead progressively
to the final property. An illustration of this process is presented in Figure 3. \\

\begin{figure}
\centering
\includegraphics[width=8cm]{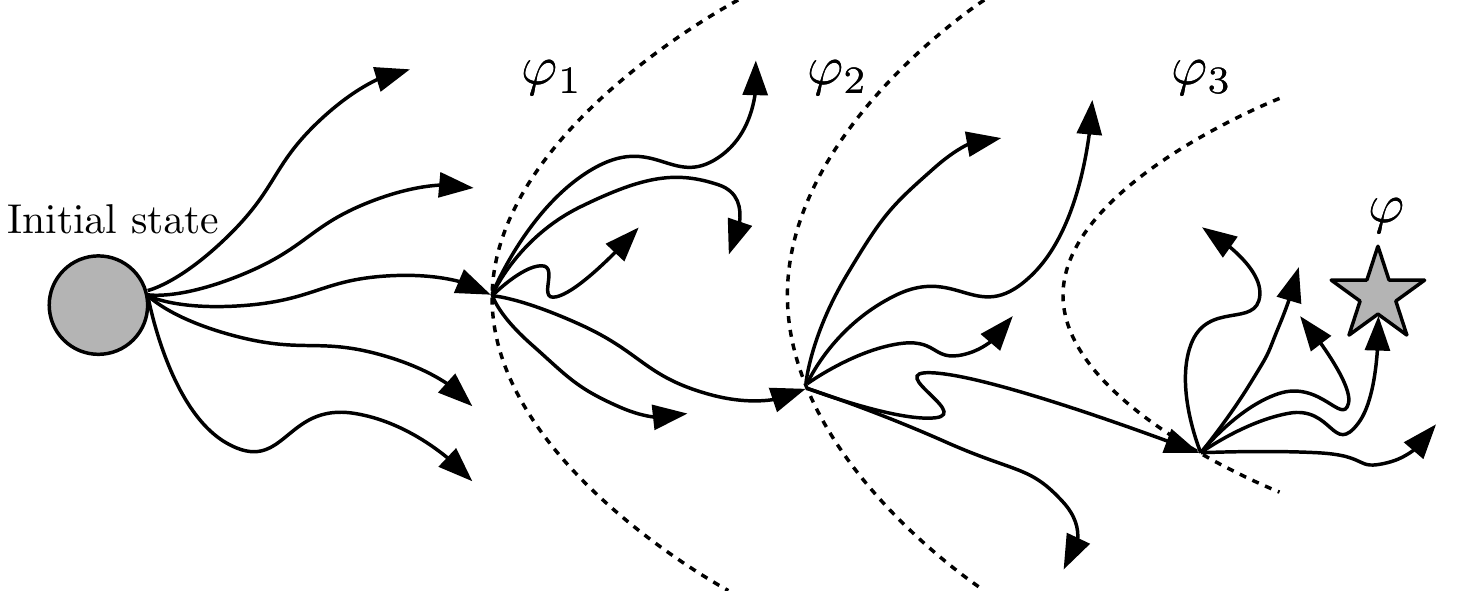}
\label{levImp}
\caption{Importance splitting using three levels}
\end{figure}

The next step is to embed the simulator, the BLTL properties and the statistical model checking algorithms into a same framework.
Plasma Lab platform is used for that.
Plasma Lab includes several statistical model checking algorithms and a library to include new simulators and to define properties.
Simulators of systems or models can be reused with few implementation work thanks to the existing libraries.
Plasma Lab already includes simulators for the Reactive Module Language (Markov chains models as in the PRISM model-checker), biological models, MATLAB/Simulink models, SystemC. In our case, we have developed a new plug-in that implements
Plasma Lab library and creates an interface between Plasma Lab and the simulator presented in the previous section.


%
\section{Experiments}
The aim of this section is to analyse the validity of our approach through experimental results.
Indeed, now that we have a model and a verification process, we need to be sure that it will efficiently detect the errors leading to safety or availability issues.
For the first experiment, we introduced several errors in the application data in order to test if they will be detected through simulations:

\begin{itemize}
\item[a.] Missing condition in a route request (removing a condition in Listing \ref{routeRequest}).
\item[b.] Point moved to a wrong position when setting a route (modifying a \texttt{cn} into a \texttt{cr}  or vice versa in Listing \ref{routeRequest}).
\item[c.] Subroute not properly locked when setting a route (removing a subroute locking in Listing \ref{routeRequest}).
\item[d.] Condition missing for releasing a subroute (removing a condition in Listing \ref{compFreeing}).
\item[e.] Condition missing for releasing an immobilisation zone (removing a condition in Listing \ref{compFreeing_UIR}).
\item[f.] Irrelevant additional conditions for releasing a component (adding a condition in  Listing \ref{compFreeing} or \ref{compFreeing_UIR}). \\
\end{itemize}

\subsubsection*{Monte Carlo estimations}

Using Monte Carlo estimations, we performed 100 simulations in order to cover 100 complete days.
Table 1 recaps the execution time and the probability to detect issues violating the requirements formalised on Section \ref{propDefSec} when errors are introduced.
Simulating the trace and verifying all the requirements for a single simulation take approximately 15 seconds. \\

\begin{table}[ht]
\centering
\caption{Execution time (in seconds) and probability (in percent) of detecting an issue violating requirements of Section \ref{propDefSec} when errors are inserted on Braine l'Alleud application data}
\begin{tabular}{|c||c|c|c||c|c||c|}
 \hline
 \multicolumn{1}{|c||}{}& \multicolumn{3}{c||}{Safety requirements} & \multicolumn{2}{c||}{Availability requirements} & \multicolumn{1}{c|}{Time (sec.)} \\
 \hline
& (1) & (2) & (3) & (4) & (5) &\\
\hline
\hline
a. & 0 &	100 &	0	& 0	& 0	& 1424 \\
\hline
b. & 0	& 0& 	100 & 	100	& 98& 	1086\\
\hline
c. & 91	& 69	& 29	& 63& 	100	& 1348\\
\hline
d. & 93 &	100	& 0	& 33	& 99	& 1845\\
\hline
e. & 72	& 97	& 0	& 79	& 100	& 1199\\
\hline
f. & 0& 0	& 0	& 100	& 100	& 1652\\
\hline
\end{tabular}

\label{errTable}

\end{table}

A non-zero probability means that a safety or availability issue occurred on at least one simulation. In this case, we have then the certainty that the interlocking is not correct.
As we can see in Table 1, each error introduced in the application causes the violation of at least one requirement 
which means that all the errors have been detected through 100 simulations of one day.
Using these results, it is possible to analyse which issues are caused by specific errors. For instance, injecting an error of type f. only causes availability issues. Such an error could not be detected in \cite{cappart_simu}.
Concerning the execution time, it remains similar in the different cases, and is consistent with the time needed to execute a single simulation. \\

\subsubsection*{Chernoff's bound}

Besides the error detection, it is also interesting to have guarantees about the correctness of the model.
Chernoff's bound is used in order to obtain a confidence interval on the probability that the requirements are satisfied in the model.
Using Equation \eqref{chernoff}, a $(1-\delta)$-$[\gamma-\epsilon,\gamma+\epsilon]$ confidence interval on $\gamma$ with $\epsilon =  0.01$ and $\delta = 0.01$ can be obtained with 26 492 simulations.
Furthermore, refining $\delta$ by a factor 10 will increase the number of required simulations by 11513 while refining $\epsilon$ by the same factor will multiply this number by 100. Knowing the execution time required to perform a 1-day simulation ($\simeq 15$ sec), we can deduce the expected time required to obtain such confidence intervals.
For $\epsilon =  0.01$ and $\delta = 0.01$, the execution will take approximately 5 days and tightening  $\delta$ by a factor 10 will add 2 days of computation while
refining $\epsilon$ by the same factor will multiply the execution time by 100. \\

Even if poor confidence intervals can be obtained in a reasonable time, more accurate intervals rapidly become too long to obtain on a single processor.
However, unlike model checking which is difficult to parallelise, the simulations can be executed in parallel without any overhead.
An execution on $m$ processors will thus divide to total execution time by $m$, which makes it possible to use in practice.
Furthermore, as shown in the first experiment, error detection requires generally far more less simulations than the number determined by Chernoff's bound. \\

\subsubsection*{Importance splitting for collision detection}

The last experiment deals with the safety requirement stating that no collision can occur in the station (Formula (1)).
As explained in the previous section, importance splitting can be used to speed up the errors detection.
Furthermore, the no collision requirement can easily be split in different levels.
For these reasons, we also employed the importance splitting algorithm for this requirement. \\

The first step is to define the different levels.
The first level is reached when two conflictual routes are set together in the station. Two routes are conflictual if they share at least one common track segment. For instance, R\_CXC\_102 and R\_DXC\_091 in Figure \ref{fig:braine} are conflictual. The next level is reached when there is only one track segment between two trains following conflictual routes if and only if no train is beyond the track segment where the collision can occur. 
According to the previous example, if the train following Route R\_CXC\_102 is on Track segment T\_01BC and the train following Route R\_DXC\_091 on Track 102, 
there is only a difference of one track segment. The third level is the event that we want to detect: the collision. According to importance splitting algorithm, simulations reaching a level are recorded and then used as a new start point for next simulations.  \\


Once the levels are determined, importance splitting can be used.
Table 2 presents statistics for simulations of 1 day with an error of type d. on the application data. Confidence intervals are obtained using a normal distribution as suggested in Section 5.2 of \cite{jegourel2014effective}. The number of experiments and the number of simulations per experiment are chosen in order to have the same total number of simulations for Monte Carlo and importance splitting (1000 in that case). As we can see, even for a medium size station such as Braine l'Alleud where errors are rapidly detected with Monte Carlo, importance splitting can give similar results much faster for a same number of simulations. 
%

\begin{table}[ht]
\centering
\caption{Comparison between Monte Carlo and importance splitting algorithms for collision detection when an error of type f. is introduced on Braine l'Alleud application data}
\begin{tabular}{|c||c|c|}
 \hline
Statistics & Importance splitting & Monte Carlo \\
\hline
\hline
\# experiments & 10    & 1 \\
\hline
\# simulations per experiment & 100  & 1000 \\
\hline
Average execution time (sec) & 257   &  1320 \\
\hline
Average arithmetic mean (\%)  & 93.9   & 93.94 \\
\hline
Standard deviation (\%) & 1.21   & 0.75 \\
\hline
$99.9\%$-confidence interval (\%)  & $[92.71,95.12]$  & $[91.43,96.45]$ \\
\hline
\end{tabular}

\label{importanceTab}

\end{table}

\section{Conclusion}

Automatic verification of an interlocking system is an active field of investigation in the railway domain.
Up to now, most of the research dealing with this issue is based on model checking or, more recently, on a discrete event simulation approach.
However, both of these approaches have drawbacks. On the one hand, model checking suffers from the state space explosion problem which complicates its use for large stations and on the other hand, discrete event simulation does not provide 
a sufficient guarantee that the system is correct.
In this paper, we propose an intermediate approach based on statistical model checking that overcomes both issues.
The key idea is to perform several simulations and to observe through hypothesis testing and other statistical tests whether the results obtained provide a statistical evidence that the system is correct.
Safety and availability properties have been formulated, and experimental results have shown that our verification approach thoroughly detects violations of the properties with a confidence interval on the verdict.
Furthermore, the importance splitting algorithm has been used in order to speed up the detection of some safety issues.\\

Up to now, the simulation performed only considers scenarios where the interlocking manages a perfect train traffic.
Unexpected events, like a train failure or a faulty sensor, are not taken into account and can be considered as rare events of a simulation. A possible improvement is to extend the simulator in order to cover such scenarios and then, to apply importance splitting in order to detect such rare events. The challenge is to adequately split the properties into consistent and relevant levels.
Our future work will be to apply our method on a larger station containing a more complex interlocking and to analyse how the verification and its performance are impacted. Station of Courtrai (70 routes, 19 track segments, 26 points and 24 signals) is considered for our next case study.

\section*{Acknowledgment}
This research is financed by the Walloon Region as part of the Logistics in Wallonia competitiveness pole.
\nocite{*}
\bibliographystyle{IEEEtran}
\bibliography{IEEEabrv,FACBib}
\end{document}